\documentclass[aps,pra,twocolumn,showpacs,10pt]{revtex4-1}

\usepackage{graphicx}
\usepackage{dcolumn}
\usepackage{bm}
\usepackage{amsmath}
\usepackage{amssymb}
\usepackage[utf8x]{inputenc}
\usepackage[T1]{fontenc}
\usepackage[%
  colorlinks=true,%
	linkcolor=blue,%
  citecolor=blue,%
  urlcolor=blue%
]{hyperref}
\newcommand{\ii}{\mathrm{i}}
\DeclareMathOperator{\ee}{e}
\DeclareMathOperator{\artanh}{artanh}
\newcommand{\dd}{\mathrm{d}}

\usepackage{txfonts}
\DeclareMathAlphabet{\mathcal}{OMS}{cmsy}{m}{n}

\begin{document}

\title{Quantum phase transitions in networks of Lipkin--Meshkov--Glick models}

\author{A. V. Sorokin}
\email{a.sorokin@mailbox.tu-berlin.de}
\author{V. M. Bastidas}
\email{victor@physik.tu-berlin.de}
\author{T. Brandes}
\affiliation{Institut für Theoretische Physik, Technische Universität Berlin, Hardenbergstr. 36, D-10623, Berlin, Germany}

\begin{abstract}
We study the quantum critical behavior of networks consisting of Lipkin--Meshkov--Glick models with an anisotropic ferromagnetic coupling. We focus on the low-energy properties of the system within a mean-field approach and the quantum corrections around the mean-field solution.  Our results show that the weak-coupling regime corresponds to the paramagnetic phase when the local field dominates the dynamics, but the local anisotropy leads to the existence of an exponentially-degenerate ground state.  In the strong-coupling regime, the ground state is twofold degenerate and possesses long-range magnetic ordering.  Analytical results for a network with the ring topology are obtained.
\end{abstract}

\pacs{05.30.-d, 05.30.Rt, 05.45.Xt, 03.65.Sq}

\maketitle

\section{Introduction}\label{sec:intro}
The Lipkin--Meshkov--Glick (LMG) model describes an ensemble of all-to-all--coupled two-level systems with anisotropic interactions \cite{1965-Lipkin-NuclPhys,*1965-Meshkov-NuclPhys,*1965-Glick-NuclPhys}.  This model is complex enough to show quantum phase transitions (QPT) subject to the change of parameters, but it is exactly solvable in the thermodynamic limit \cite{2007-Ribeiro-PRL}.  The total angular momentum that is formed by combining all the spins of single particles is, in this limit, so long that its behavior is close to classical, and the properties of the system can be rather precisely described in the mean-field approximation.  From a theoretical point of view, there are proposals to realize the LMG model by means of cavity QED setups \cite{2008-Morrison-PRA,*2008-Morrison-PRL}. Further theoretical approaches have shown intriguing relations to quantum Fisher information \cite{2009-Ma-PRA}, and spin squeezing \cite{1993-Kitagawa-PRA, 2011-Ma-PhysRep}.  Experimentally, the dynamics of LMG model has been explored by using Bose--Einstein condensates \cite{2010-Gross-Nature,2010-Zibold-PRL,2006-Morsch-RevModPhys,2010-Theocharis-PRA}. 

Networks of coupled critical systems like LMG models may show new phases with different long- and short-range ordering depending on the topology of the network. In this paper, we use Holstein--Primakoff transformations \cite{1940-Holstein-PhysRev} and a mean-field approach to describe quantum phase transitions in a network composed of LMG models with anisotropic ferromagnetic interactions between different sites.  Altogether, this restricts us to the lowest energy states only and allows us to study the quantum fluctuations about the mean field.  Working in low-energy regions is also the reason for not experiencing any chaotic behavior in the semiclassical limit, even for networks with more than two degrees of freedom \cite{2011-Chotorlishvili-PRB}.

Related works used Holstein--Primakoff transformations to describe low-energy magnetic excitations in time-dependent magnetic fields \cite{1971-Zagury-PRB} and the interaction of magnons in Heisenberg ferromagnets \cite{2010-Cheng-PhysicaA}. Furthermore, in the context of spinor Bose--Einstein condensates \cite{2013-Stamper-RevModPhys,2014-Marti-arXiv}, Holstein--Primakoff transformations can be used to describe the formation of periodic magnetic domains \cite{2006-Zheng-AnnPhys}.  In most of the papers concerning long-spin chains, the coupling is chosen to have certain continuous symmetries, most commonly by using either isotropic Heisenberg-type or anisotropic coupling \cite{2011-Chotorlishvili-PRB,2011-Konstantinidis-JPhys,2011-Karimi-PRB,2006-Zheng-AnnPhys}.  In this paper, however, we focus on the uniaxial coupling, which leads to the existence of a set of global and local discrete symmetries.  Thus, the absence of rotational symmetry and the emergence of local discrete symmetries in the problem open the possibility of new effects. Previous works explored dynamical aspects of networks of coupled systems with global symmetries.  For example, the adiabatic phase transitions of networks of qubits were investigated \cite{2014-Acevedo-PRL}. In the context of quantum optics, arrays of coupled cavities can exhibit soliton solutions \cite{2012-Chen-PRA}, the emergence of phase transitions of light \cite{2012-Schiro-PRL}, and dissipative quantum phase transitions \cite{2011-Liu-PRA}.

The intriguing properties of spin networks with spatial symmetries have found many experimental implementations.  For instance, the chains of trapped ions were shown to undergo a variety of quantum phase transitions when interacting with the laser beams \cite{2004-Porras-PRL}.  They were as well used to detect quantum correlations between a two-level system and the environment by measuring the system only \cite{2014-Gessner-NatPhys}.  Other experimental implementations of critical spin chains include ultracold polar molecules \cite{2013-Yan-Nature} and Rydberg gases \cite{2012-Schauss-Nature} to name but a few. 

The structure of the paper is as follows.
In Sec.~\ref{sec:model}, we describe the model and our bosonization approach.  The latter is then used to calculate the ground-state energy analytically in the thermodynamic limit and to identify the different phases of the system. We compare our analytical results with the exact diagonalization of the Hamiltonian in the case of the finite total angular momentum $j$ and the small number of sites $N$ in the chain. 
In Sec.~\ref{sec:dispersion}, we calculate dispersion relations for the excitation energy in different phases by using Bogoliubov and discrete Fourier transformations.  The behavior of the low-energy excitations near phase boundaries is then discussed.
In Sec.~\ref{sec:correlation}, we calculate correlation functions in the ground state with full translational invariance.  

\section{The model}\label{sec:model}
In this paper we consider a set of coupled LMG models 
\begin{equation}
  \mathcal{H}_l = g J^{z}_l-\frac{\gamma}{2j}(J_l^x)^{2},
  \label{eq:1N-LMG}
\end{equation}
each of which is represented by a node in a network.  Throughout the text, $g$ is the strength of an external field
and $\gamma$ determines the self-interaction. We define the $\xi$\nobreakdash-component of the collective angular momentum at the $l$\nobreakdash-th site $J_l^{\xi}=\frac{1}{2}\sum_{a=1}^n{\sigma^{\xi}_{al}}$, where $\sigma^{\xi}_{al}$ are Pauli matrices satisfying the algebra $[\sigma^{\xi}_{al},\sigma^{\nu}_{bl'}]=2\ii\varepsilon^{\xi\nu\rho}\delta_{a b}\delta_{l l'}\sigma^{\rho}_{al}$ and the indices $\xi,\nu,\rho \in \{x,y,z\}$ denote the spin components. For a fixed length of the collective angular momentum $j$, the Hamiltonian \eqref{eq:1N-LMG} undergoes a second-order QPT at $\gamma=g$ \cite{2007-Ribeiro-PRL}. 

In this paper, we study the critical behavior of networks of quantum critical systems by assuming nondirected coupling between the $J_y$ components of the nodes, so the Hamiltonian reads
\begin{equation}
	\mathcal{H}=\sum_{l=1}^N \mathcal{H}_l - \frac{1}{2j}\sum_{l'\geqslant l=1}^N\kappa_{ll'}J_l^y J_{l'}^y,
  \label{eq:NN-LMG}
\end{equation}
where $\kappa_{ll'}$ is the coupling matrix of the network \cite{2008-Arenas-PhysRep,1985-Chartrand} and $l,l' \in \{1,2,\ldots,N\}$ denote the sites of the chain.  The collective angular momentum operators satisfy commutation relations $[J_l^{\xi},J_{l'}^{\nu}]=\ii\varepsilon^{\xi\nu\rho}\delta_{l l'}J_l^{\rho}$.  In the model, we assume the constants $g$ and $\gamma$ to be positive, and by choosing $\kappa_{ll'}\geqslant 0$, we restrict ourselves to ferromagnetic coupling.  The model we are considering is a minimal example of critical networks, so we shall not look at the antiferromagnetic case $\kappa_{ll'}\leqslant 0$ here, as that would ask for facing the intricacies of frustration \cite{2010-Balents-Nature}.

\begin{figure}
  \includegraphics[width=0.8\columnwidth]{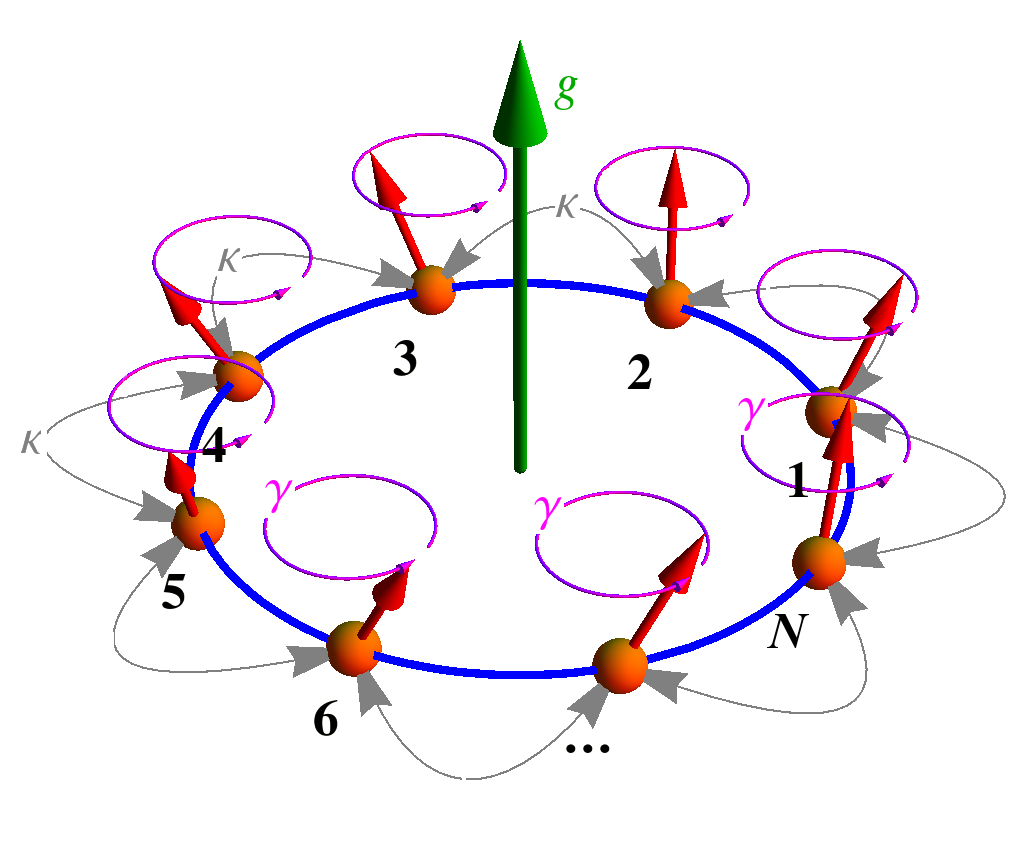}
  \caption{\label{fig:NN-Ring}(Color online) Ring network of LMG models.  The coupling between neighboring sites is determined by parameter $\kappa$, the strength of interaction within a single site by $\gamma$, and $g$ models an external field.}
\end{figure}

Up to Sec.~\ref{sec:finiteN}, the actual topology of the network is irrelevant, but afterwards it is set to the ring-type one as shown in Fig.~\ref{fig:NN-Ring} by additionally implying periodic boundary conditions $J^\xi_{N+1}\equiv J^\xi_1$.  This type of network introduces additional translational symmetry that allows for Fourier transformations and thus simplifies the calculations.

\subsection{Symmetries and limit cases}\label{sec:symmetries}
The Hamiltonian \eqref{eq:NN-LMG} preserves the local angular momentum $[\mathcal{H},\bm{J}^{2}_{l}]=0$, where $\bm{J}_l^2=(J^x_l)^2+(J^y_l)^2+(J^z_l)^2$.  Therefore, we can fix $j$ to its maximal value $n/2$ throughout the paper.  This implies that instead of working in a Hilbert space with the dimension $d=2^{nN}$, we can restrict ourselves to a subspace with the dimension $d_s=(n+1)^{N}$ spanned by the basis of tensor products of Dicke states of the individual nodes
\begin{equation}
  |j,\bm{m}\rangle^\xi=\bigotimes_{l=1}^N |j,m_l\rangle^\xi,
\end{equation}
where $\bm{m}=(m_1,m_2,\ldots,m_N)$, $-j\leqslant m_l\leqslant j$, and $\xi$ denotes the quantization axis.  The states $|j,m_l\rangle^\xi$ are eigenstates of the collective angular momentum operators $J^\xi_l$, such that $J^\xi_l|j,m_l\rangle^\xi=m_l|j,m_l\rangle^\xi$.

The Hamiltonian \eqref{eq:NN-LMG} possesses the global parity
\begin{equation}
  \Pi=\exp\left[\ii\pi\sum_{l=1}^N\left(J_l^z+j\right)\right],
  \label{eq:GlobalParity}
\end{equation}
which is just a product of parities of individual nodes \cite{2003-Emary-PRL,2007-Ribeiro-PRL}.  Under the action of $\Pi$, the total angular momentum transforms as $\Pi\,(J_l^x, J_l^y, J_l^z)\,\Pi^\dagger=(-J_l^x, -J_l^y, J_l^z)$.  In the next sections, most importantly in order to perform numerical calculations efficiently, we construct the basis from tensor products of eigenstates of $J^z_l$.  The global parity operator \eqref{eq:GlobalParity} acts on these basis states as $\Pi\,|j,\bm{m}\rangle^z=(-1)^{\sum_l (m_l+j)}|j,\bm{m}\rangle^z$, allowing us to separate the Hilbert space into two subspaces---with positive and with negative parity.  The positive-parity subspace contains the ground state and has the dimension $d_{s+}=(n+1)^N/2$.

Apart from the global parity \eqref{eq:GlobalParity}, the system is also invariant under the local reflection $\mathcal{R}^{yz}_l$ in the $yz$\nobreakdash-plane
\begin{align}    
      \mathcal{R}^{yz}_{l} &= \exp\left[\ii\pi\left(J_l^x+j\right)\right]\exp\left(\ii\pi J_l^y\right)\mathcal{K}_l\nonumber \\
                           &= \exp\left[\ii\pi\left(J_l^x+j\right)\right]\mathcal{T}_l,
      \label{eq:LocalReflection}
\end{align}
where $\mathcal{T}_l=\exp\left(\ii\pi J_l^y\right)\mathcal{K}_l$ is the time-reversal operator and $\mathcal{K}_l$ is the operator of charge complex conjugation with respect to the standard representation \cite{2010-Haake} acting on the $l$\nobreakdash-th site.  The action of the anti-unitary local reflection operator on the angular momentum reads $\mathcal{R}^{yz}_{l}\,(J_l^x, J_l^y, J_l^z)\,(\mathcal{R}^{yz}_{l})^{-1}=(-J_l^x, J_l^y, J_l^z)$.

Now we focus on the analysis of the limit cases to understand the properties of the ground state.  For convenience, we introduce the states
\begin{equation}
      |G^\xi_{p_1,p_2,\dots,p_N}\rangle=\bigotimes_{l=1}^N |j,(-1)^{p_l}j\rangle^\xi,
      \label{eq:GStates}
\end{equation}
where $p_{i}\in \{0,1\}$ and $\xi \in \{x,y,z\}$.

In the limit $g\gg\gamma,\kappa_{ll'}$, there is a unique ground state $|G\rangle=|G^z_{1,1,\dots,1}\rangle$---a paramagnetic-like state with short-range correlations (cf. Ref.~\onlinecite{2011-Sachdev}).

In the limit $\gamma \gg g, \kappa_{ll'}$ the ground state is $2^N$-fold degenerate and is represented by the set of separable states $|G^x_{p_1,p_2,\dots,p_N}\rangle$ with all the possible combinations of $p_i$.  In this regime, the system consists of an ensemble of $n$ tightly-bound particles with parallel spins along the $x$\nobreakdash-direction at each site of the network.  The exponential degeneracy of the ground state in this regime is a consequence of the local symmetry \eqref{eq:LocalReflection}. It is worth noting that exponentially-degenerate ground states arise naturally in the context of spin ice \cite{2010-Balents-Nature} and spin glasses \cite{1986-Binder-RevModPhys}.

Finally, in the strong interaction limit $\kappa_{ll'} \gg g, \gamma$, the ground state is highly correlated, twofold degenerate, and includes ferromagnetic states $|G^y_{0,0,\dots,0}\rangle$ and $|G^y_{1,1,\dots,1}\rangle$.

From the analysis of limit cases we can conclude that the ground states in different limits are drastically different, so the properties between these limits should behave nonanalytically at some points.  This is the onset of the critical behavior that we seek to describe in this paper.

\subsection{Bosonization and the ground-state energy}
As we are working in the thermodynamic limit, i.e. $j$ is sufficiently large, we can map the angular momentum operators $\bm{J}_l=(J_l^x, J_l^y, J_l^z)$ onto bosonic operators $b_l$, $b_l^\dagger$, which satisfy the commutation relations $[b_l, b_{l'}^\dagger]=\delta_{ll'}$ and $[b_l, b_{l'}]=0$, using Holstein--Primakoff transformations \cite{1940-Holstein-PhysRev}
\begin{gather}
	J_l^z = b_l^\dagger b_l-j,\nonumber\\
	J_l^+ = b_l^\dagger\sqrt{2j-b_l^\dagger b_l},\qquad
	J_l^- = \sqrt{2j-b_l^\dagger b_l}\;b_l.
	\label{eq:HP-OpsO}
\end{gather}
With these transformations, the harmonic approximation around a fixed point is done.  In order to obtain the mean-field configurations, we replace the original operators $b_l$ with displaced operators:
\begin{equation}
  b_l=\mathcal{D}^\dagger\left(\alpha_l\sqrt{j}\right)\,d_l\,\mathcal{D}\left(\alpha_l\sqrt{j}\right)=d_l+\alpha_l\sqrt{j},
  \label{eq:HP-Ops}
\end{equation}
where $\alpha_l$ are the mean fields for each of the nodes, $d_l$ are quantum fluctuations around these, and we define the bosonic displacement operator \cite{1969-Glauber-PhysRev}
\begin{equation}
  \mathcal{D}\left(\alpha_l\sqrt{j}\right)=\exp\left[\left(\alpha_l d_l^{\dagger}-\alpha_l^* d_l\right)\sqrt{j}\right].
  \label{eq:DisplacementOperator}
\end{equation}

By substituting \eqref{eq:HP-Ops} into \eqref{eq:HP-OpsO} and expanding the radicals up to $O[(b_l^\dagger b_l/2j)^2]$ as in Refs.~\onlinecite{2003-Emary-PRE,2003-Emary-PRL}, we get angular momentum operators expressed in terms of $\alpha_l$ and $d_l$:
\begin{gather}
	J_l^z = d_l^\dagger d_l + \left(\alpha_l^*d_l+\alpha_l d_l^\dagger\right)\sqrt{j} + \left(|\alpha_l|^2-1\right)j,\nonumber\\
	J_l^+ = \sqrt{j\left(2-|\alpha_l|^2\right)}\left(d_l^\dagger+\alpha_l^*\sqrt{j}\right)\left(1-\frac{c_l}{2}-\frac{c_l^2}{8}\right),\label{eq:HP-OpsOE}\\
	J_l^- = \sqrt{j\left(2-|\alpha_l|^2\right)}\left(1-\frac{c_l}{2}-\frac{c_l^2}{8}\right)\left(d_l+\alpha_l\sqrt{j}\right)\nonumber
\end{gather}
with
\[c_l=\frac{d_l^\dagger d_l + (\alpha_l^*d_l+\alpha_l d_l^\dagger)\sqrt{j}}{(2-|\alpha_l|^2)j}.\]

Substituting \eqref{eq:HP-OpsOE} into \eqref{eq:NN-LMG} and truncating higher-order terms we reduce the Hamiltonian to the form
\begin{equation}
	\mathcal{H}=E_g(\bm{\alpha})j+\mathcal{H}_L(\bm{d},\bm{\alpha})\sqrt{j}+\mathcal{H}_Q(\bm{d},\bm{\alpha})
	\label{eq:NN-H-QLG}
\end{equation}
with $\bm{\alpha}=(\alpha_1,\alpha_2,\ldots,\alpha_N)$ and $\bm{d}=(d_1,d_2,\ldots,d_N)$.  Terms $\mathcal{H}_L$ and $\mathcal{H}_Q$ are, respectively, linear and quadratic in bosonic operators \cite{2011-Hayn-PRA,*2012-Hayn-PRA}.

\begin{figure}
	\includegraphics[width=0.8\columnwidth]{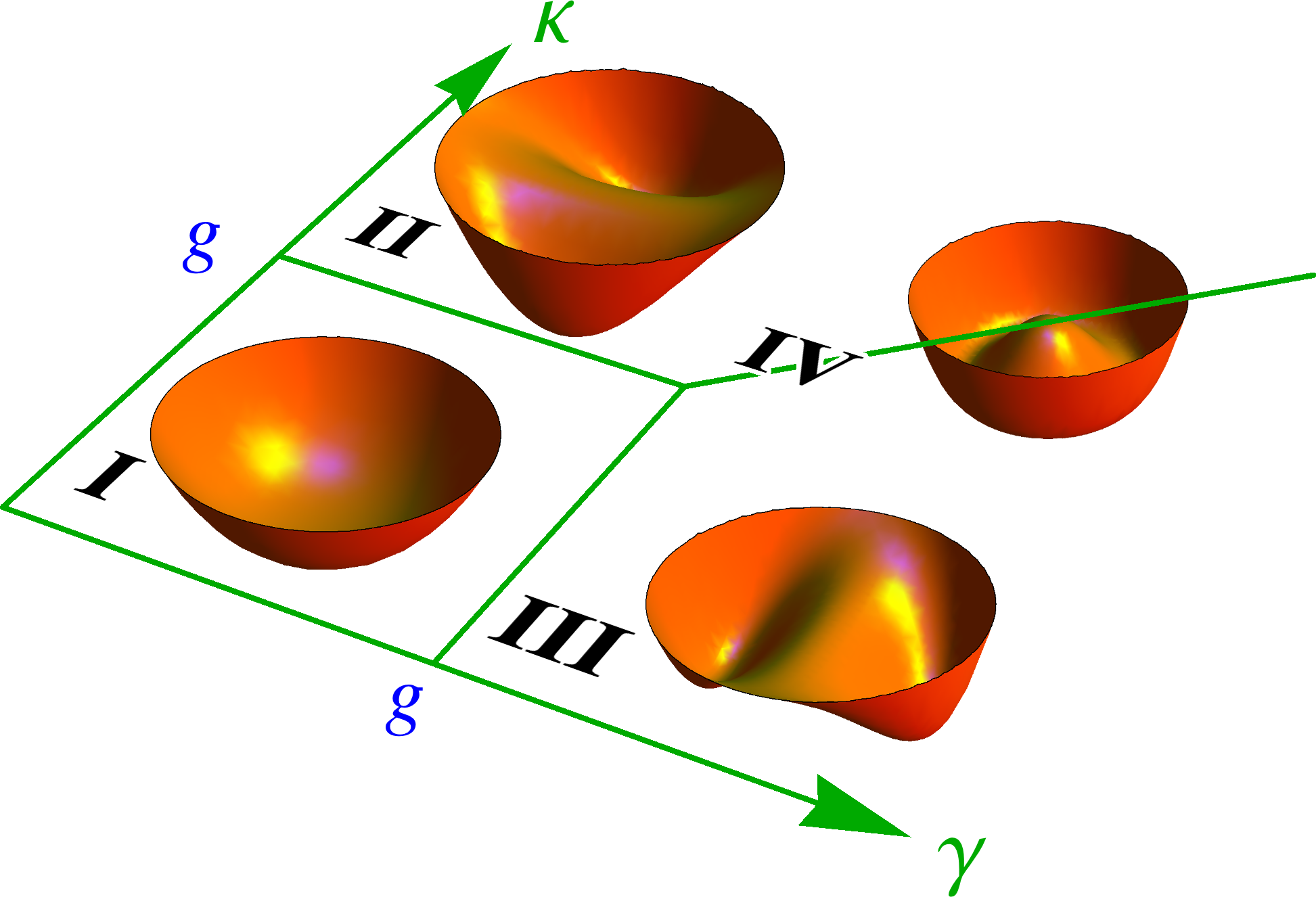}
	\caption{\label{fig:NN-Phases}(Color online) The phase diagram for the ring network of LMG models.
	The classical energy surfaces $E_g(\alpha)$ are shown in each of the characteristic regions.}
\end{figure}

The $O(j)$ terms in the expansion of the transformed Hamiltonian add up to form the ground-state energy of the system, which depends on mean fields $\alpha_i$ of each of the nodes and takes the form
\begin{align}
	E_g(\bm\alpha) =&\;
		Ng -\frac{\gamma}{4}\sum_l\left(\alpha _l^2+\alpha _l^{*2}\right)-\left(\frac{\gamma}{2}+g\right) \sum_l\alpha _l{}^* \alpha _l \nonumber\\
            &+\frac{\gamma }{8}\sum_l \alpha_l^*\alpha_l\left(\alpha_l+\alpha_l^*\right)^2 \nonumber\\
		&+\frac18\sum_{l'\geqslant l}\kappa_{ll'}\bigg[\left(\alpha_l-\alpha_l^*\right) \left(\alpha_{l'}-\alpha_{l'}^*\right) \nonumber\\
		&\quad\qquad\times\sqrt{2-\alpha_l^* \alpha_l} \sqrt{2-\alpha_{l'}^* \alpha_{l'}} \bigg].
	\label{eq:NN-EG}
\end{align}
We are altogether interested in such $\bm{\alpha}$ values that would minimize $E_g(\bm\alpha)$, as these would correspond to the stable fixed points of the network.  The solution of $2N$ simultaneous equations $\{\partial_{\alpha_i} E_g(\bm\alpha)=0,\; \partial_{\alpha_i^*} E_g(\bm\alpha)=0\}$, which would give us all the critical points of the surface cannot be obtained analytically even for $N$ as low as 2.  That leaves us with the necessity of locating the critical points numerically.

Another approach we can take---justified both by numerical diagonalization and by symmetry reasons---is to assume that the global minimum (or at least one of the global minima, if they are degenerate) is located at the points of identical mean fields (cf. Ref.~\onlinecite{2009-Vidal-PRB}).  In this case, the ground-state energy becomes a function of only one complex variable $\alpha=\alpha_1=\ldots=\alpha_N$, and the expression \eqref{eq:NN-EG} simplifies to
\begin{align}
	E_g(\alpha) =&\;
		Ng -N\frac{\gamma}{4}\left(\alpha^2+\alpha^{*2}\right)-N\left(\frac{\gamma}{2}+g\right) \alpha^* \alpha \nonumber\\
		&+N\frac{\gamma }{8}\alpha^*\alpha\left(\alpha_l+\alpha_l^*\right)^2 \nonumber\\
            &+\frac18\left(\alpha-\alpha^*\right)^2 \left(2-\alpha^* \alpha\right)\sum_{l'\geqslant l}\kappa_{ll'}.
	\label{eq:NN-EGa}
\end{align}
The simultaneous equations $\{\partial_{\alpha} E_g(\alpha)=0,\; \partial_{\alpha^*} E_g(\alpha)=0\}$ can now be solved analytically for an arbitrary network.

As in the next sections we shall explore one specific network type, namely the looped chain (see Fig.~\ref{fig:NN-Ring}), we look for the solutions of these equations with $\sum\kappa_{ll'}=N\kappa$.  Taking into account the constraint $\alpha^* \alpha\leqslant 2$ dictated by the reality of the roots in \eqref{eq:NN-EG}, the only possible critical points in this network are
\begin{equation}
  \alpha_g=0,\quad \alpha_{\gamma\pm}=\pm\sqrt{1-\frac{g}{\gamma}},\quad \alpha_{\kappa\pm}=\pm\ii\sqrt{1-\frac{g}{\kappa}.}
  \label{eq:GroundStateMin}
\end{equation}
Of these five points, $\alpha_{\gamma\pm}$ exist only in the $\gamma$-dominated region (III), $\alpha_{\kappa\pm}$ only in the $\kappa$-dominated region (II), and $\alpha_g$ is the minimum point only in the $g$-dominated region (I) (see Fig.~\ref{fig:NN-Phases} for labels of phases).

The variations of the energy per site $E_g(\alpha)/N$ in $\gamma\kappa$\nobreakdash-space are shown in the left column of Fig.~\ref{fig:3N-EG} with thick red lines, by assuming the equality of all the mean fields as in \eqref{eq:NN-EGa}.  Fig.~\ref{fig:3N-EG} also depicts the exact numerical results for a network with the ring topology and the finite number of sites $N$.  As implied by \eqref{eq:NN-EGa}, $E_g/N$ is independent of the number of sites and thus remains the same even in the limit $N\to\infty$.  Besides, in regions (II) and (III), $E_g$ depends exclusively on one parameter---$\kappa$ or $\gamma$, respectively---while in region (I) it is constant.

\subsection{Finite $N$ case}\label{sec:finiteN}
\begin{figure}
	\includegraphics[width=\columnwidth]{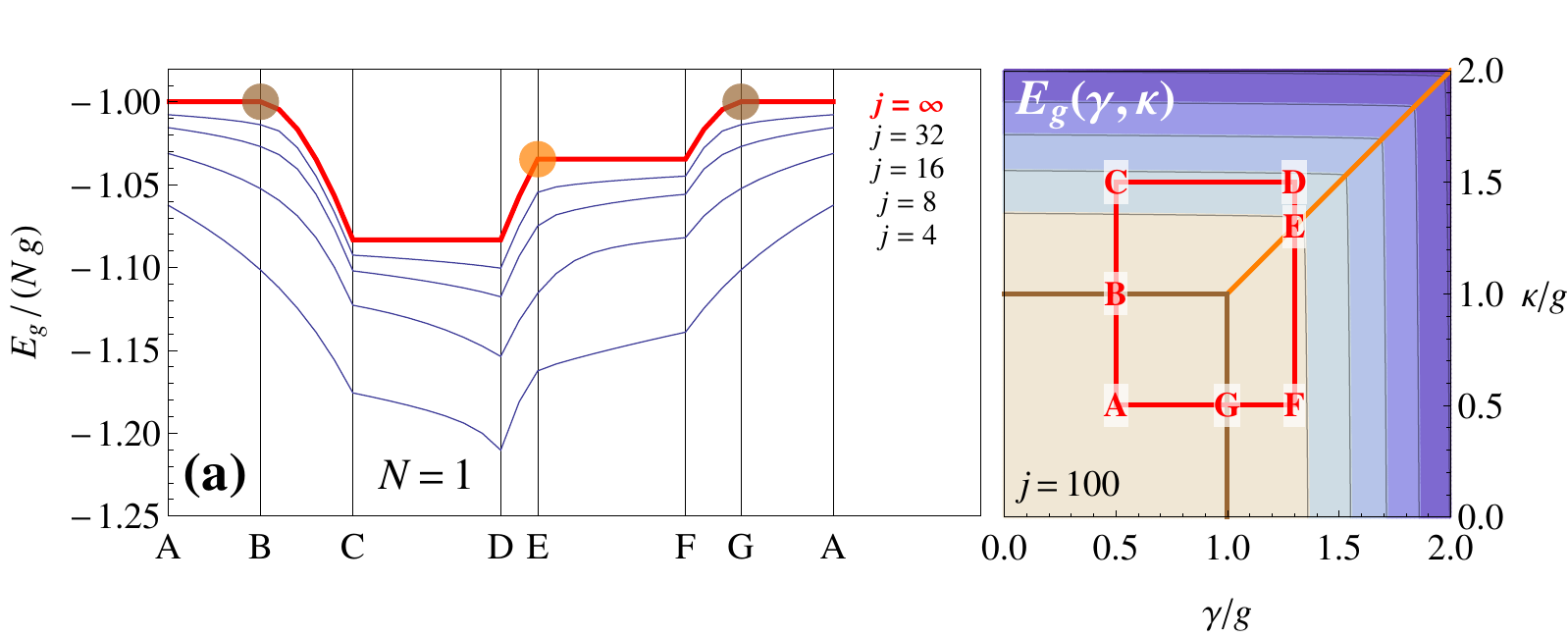}\vspace{-1ex}
	\includegraphics[width=\columnwidth]{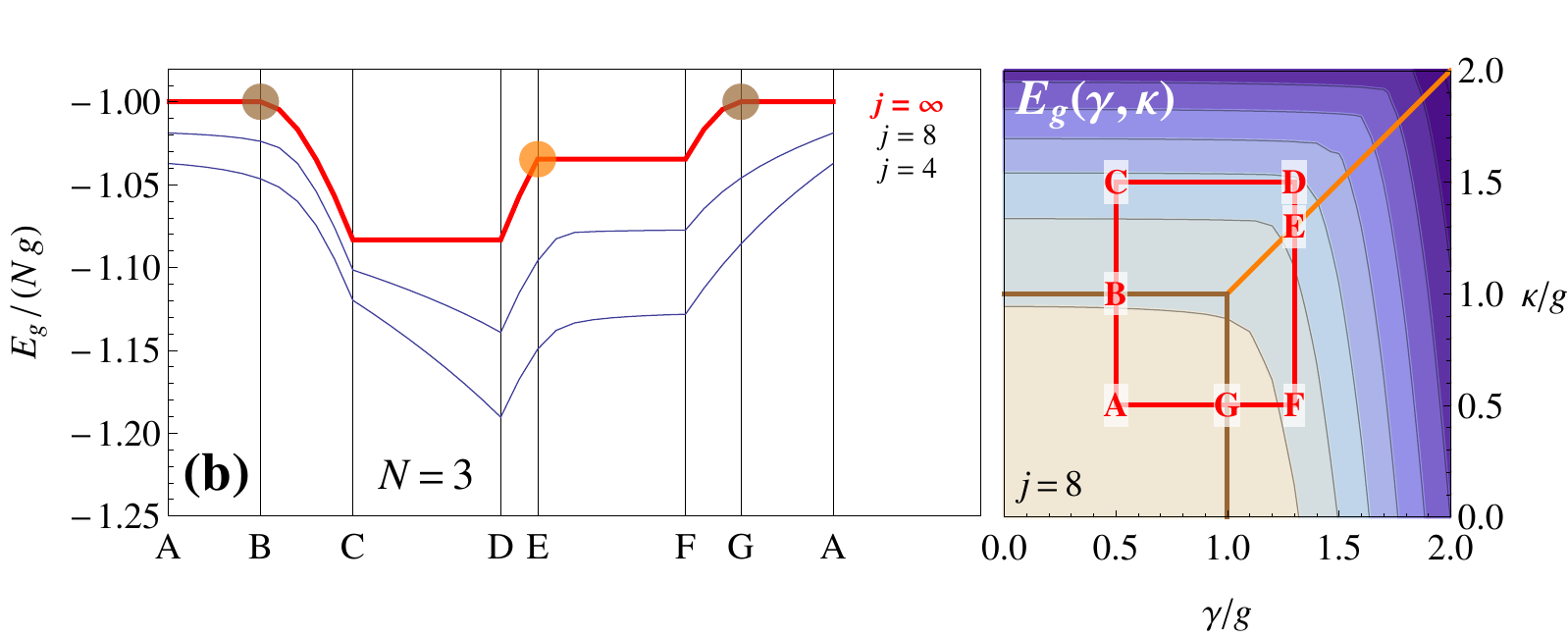}
  \includegraphics[width=\columnwidth]{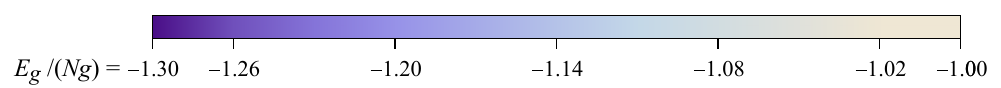}
	\caption{\label{fig:3N-EG}(Color online) Numerically calculated ground-state energy for the LMG chain with the number of sites $N=1$~(a) and $N=3$~(b) and different single-node total angular momentum values (thin lines) along the path ABCDEFGA in $\gamma\kappa$-space (see contour plots) as well as the expected thermodynamic limits (thick red lines).  Contour plots show the ground-state energy for different $\gamma$ and $\kappa$ values.  Orange (light) lines and circles denote the first-order QPT, while brown (dark) ones denote the second-order QPT.}
\end{figure}

The ansatz we adopted in the previous section, namely that the mean fields for all the nodes in the network are identical in the ground state is rather strong and its acceptability needs serious justification.  To this end, we performed a direct diagonalization of the Hamiltonian for a network of a finite number of sites $N$ with the topology of a ring by assuming $J^\xi_{N+1}\equiv J^\xi_1$ and $\kappa_{ll'}=\kappa\neq 0$ only when $l'=l\pm 1$.  Ground-state energy dependence on parameters $\gamma$ and $\kappa$ is shown in Fig.~\ref{fig:3N-EG} for $N=1$ (single LMG) and $N=3$ (simplest ring).  The plots for finite angular momentum within each of the nodes ranging from $j=4$ up to 32 for $N=1$ or up to 8 for $N=3$ (thin purple lines) clearly converge to the expected thermodynamic limit acquired using the ansatz from the previous section (thick red lines).

Calculations confirm that the phase diagram of the Hamiltonian \eqref{eq:NN-LMG} is similar to the one of a single anisotropic LMG model with $\gamma_{x}=\gamma$ and $\gamma_{y}=\kappa$ (cf. Ref.~\onlinecite{2007-Ribeiro-PRL} and Fig.~\ref{fig:3N-EG}a), though the physical meaning of the phases is strikingly different.  There exist three distinct regions in the phase space (Fig.~\ref{fig:NN-Phases}): a ``symmetric'' $g$-dominant phase (I) and two ``symmetry-broken'' $\kappa$- and $\gamma$-dominant phases (II) and (III), respectively.  In the phase (I), there is only one ground-state energy minimum at $\alpha_1=\ldots=\alpha_{N}=\alpha_g$.  Due to spontaneous symmetry breaking at the phase boundary, there appear two distinct ground-state energy minima at $\alpha_1=\ldots=\alpha_{N}=\alpha_{\kappa\pm}$ in the phase (II).  In the phase (III), though, $E_g$ is minimized not exclusively for the state with identical mean fields: $\alpha_l$ take either of the two values $\alpha_{\gamma\pm}$ given in \eqref{eq:GroundStateMin}, thus making the ground state $2^N$-fold degenerate.  The degree of degeneracy of the ground state in different phases found numerically coincides with what was expected from the point of view of symmetries in the limit cases in Sec.~\ref{sec:symmetries}.

At the critical lines $(\gamma=1, 0\leqslant\kappa<1)$, $(0\leqslant\gamma<1, \kappa=1)$, and $(\gamma\geqslant 1, \kappa=\gamma)$ the ground state energy landscape $E_g(\alpha)$ exhibits a bifurcation, which is a signature of the quantum phase transition \cite{2003-Emary-PRL,2007-Ribeiro-PRL,2011-Hayn-PRA,*2012-Hayn-PRA}.  Besides, as can be concluded from Fig.~\ref{fig:3N-EG}, derivatives of the ground-state energy in the thermodynamic limit with respect to the parameters of the system, show no jumps at these lines (see points B and G), allowing us to classify the QPTs occurring here as of the second order.  At the line $(\kappa=\gamma>1)$, a QPT also occurs (see point E), but this time the derivative is not continuous and the transition is of the first order.

If we adhere to the identical-mean-field ansatz, then in the thermodynamic limit the coupling term in the Hamiltonian \eqref{eq:NN-LMG} effectively induces a self-interaction in $J_y$ components, reducing $\mathcal{H}$ to the Hamiltonian of a single $xy$-anisotropic LMG model on the level of the ground-state energy.  The difference in the ground-state properties of a network and a single system appears then only in the degeneracy of the ground state in the phases (II) and (III).

For further analysis it should be noted that, be the ground state degenerate or nondegenerate, the state with identical mean fields $\alpha_1=\ldots=\alpha_N=\alpha_\mathrm{cr}$ with $\alpha_\mathrm{cr}\in\{\alpha_g,\alpha_{\gamma\pm},\alpha_{\kappa\pm}\}$ depending on the phase is always a ground state, and will be used in the following sections as such.

For the calculations of the energy dispersion this does not impose any additional restrictions in the phase (III), where the ground state is highly degenerate, for the following reasons.  Dispersion relations are determined by the quadratic part of the Hamiltonian \eqref{eq:NN-H-QLG}, which can be written as $H_Q=\bm{d}^\dagger\,\mathbb{H}\,\bm{d}$, where $(\mathbb{H})_{ll'}=\partial_{\alpha_l\alpha_{l'}} E_g(\bm{\alpha})$ is the Hessian matrix for $E_g(\bm{\alpha})$.  In the phase (III), critical points $\alpha_l=\alpha_{\gamma\pm}$ are real, making $E_g$ [see \eqref{eq:NN-EG}] an even function in each of the variables $\alpha_l$ due to the local reflection symmetry \eqref{eq:LocalReflection}.  But this means that the second derivatives are even too, making the Hessian and thus $\mathcal{H}_Q$ independent of the choice of the ground state.

\section{Energy dispersion}\label{sec:dispersion}
In this section we shall focus on the lowest excitation energies of our LMG ring model. As it was justified earlier, the identical-mean-field state with $\alpha_l=\alpha_\mathrm{cr}$ is a ground state of such a ring and will be used throughout this section.  The positions of critical points were determined earlier in \eqref{eq:GroundStateMin}.

\begin{table*}
	\caption{\label{tab:L}Ground-state energy $jE_g$ and parameters $L_0$ through $L_3$ used in \eqref{eq:NN-HQ} and further on.  The ring is initialized in its ground state in the respective region with mean fields $\alpha_l=\alpha_\mathrm{cr}$.}
	\begingroup
	\setlength\arraycolsep{8pt}
	\[\begin{array}{ccccccc}
		\hline\hline
		\text{Region} & \alpha_\mathrm{cr} & jE_g & L_0 & L_1 & L_2 & L_3 \\
            \hline
		\text{I} & 0 & -N g j & -\frac12 N \gamma & g-\frac12 \gamma & -\frac14 \gamma & \frac14 \kappa \\
            \text{II} & \pm\ii\sqrt{1-g/\kappa} & -\frac{1}{2\kappa} N j (g^2+\kappa^2) & -\frac18 N \left[2\gamma\frac{g+\kappa}{\kappa}-\frac{(\kappa-g)^2}{g+\kappa}\right] & \frac14 \left[4\kappa-2\gamma+\frac{\gamma(\kappa-g)}{\kappa}+\frac{(\kappa-g)^2}{g+\kappa}\right] & -\frac18 \frac{(g^2+2g\kappa)(\gamma-\kappa)+\kappa^2(\gamma+3\kappa)}{\kappa(\kappa+g)} & \frac12 \frac{g^2}{\kappa+g} \\
		\text{III} & \pm\sqrt{1-g/\gamma} & -\frac{1}{2\gamma} N j (g^2+\gamma^2) & \frac14 N (\gamma-3g) & \frac14 (5\gamma-3g) & \frac18 (3\gamma-5g) & \frac18 \frac{\gamma+g}{\gamma} \kappa \\
		\hline\hline
	\end{array}\]
	\endgroup
\end{table*}

The ansatz about the ground state that was made allows us to calculate energy dispersion relations analytically for an arbitrary large number of sites $N$ in the chain.  For this purpose we consider the quadratic part of the Holstein--Primakoff-transformed Hamiltonian \eqref{eq:NN-H-QLG}, which reads
\begin{align}
	\mathcal{H}_Q &= L_0 + L_1\sum_l{d_l^\dagger d_l} + L_2\sum_l{\left(d_l^2 + d_l^{\dagger 2}\right)} \nonumber\\
	&\quad + L_3\sum_l{\left(d_l^\dagger - d_l\right)\left(d_{l+1}^\dagger - d_{l+1}\right)}
	\label{eq:NN-HQ}
\end{align}
with $L_0$ through $L_3$ being factors depending on parameters of the system and on the critical point in use.  For expressions determining these factors see Tab.~\ref{tab:L}.  In order to get rid of nonlocal terms, we map $\mathcal{H}_Q$ onto the reciprocal space using Fourier transformations
\begin{equation}
	D_k=\frac{1}{\sqrt{N}}\sum_{l=1}^N{d_l \ee^{-\ii k l}}
	\label{eq:DFT}
\end{equation}
to obtain the $\mathcal{H}_Q$ in terms of Fourier images of $d_l$:
\begin{align}
	\mathcal{H}_Q &= L_0 + L_1\sum_k{D_k^\dagger D_k} + L_2\sum_k{\left(D_k D_{-k}+D_k^\dagger D_{-k}^\dagger\right)} \nonumber\\
	&\quad + L_3\sum_k{\left(D_k D_{-k}\ee^{-\ii k} + D_k^\dagger D_{-k}^\dagger\ee^{\ii k} -2 D_k^\dagger D_k \cos{k}\right)},
	\label{eq:NN-HQF}
\end{align}
with $k=0, 1\frac{2\pi}{N}, \ldots,(N-1)\frac{2\pi}{N}$ and $-k\equiv N-k$.  Fourier transformations preserve commutation relations between bosonic operators, so $[D_k, D_{k'}^\dagger]=\delta_{kk'}$ and $[D_k, D_{k'}]=0$.  In order to simplify this expression, we restrict the sums to positive wavenumbers, thus getting rid of complex exponents:
\begin{align}
	\mathcal{H}_Q &= L_0 + \sum_{k>0}{\left(D_k^\dagger D_k + D_{-k}^\dagger D_{-k}\right)\left(L_1-2L_3\cos k\right)} \nonumber\\
	&\quad + \sum_{k>0}{\left(D_k D_{-k}+D_k^\dagger D_{-k}^\dagger\right)\left(2L_2+2L_3\cos k\right)}.
	\label{eq:NN-HQFP}
\end{align}
Then the quadratic part of the Hamiltonian is readily diagonalized by means of Bogoliubov transformations \cite{2008-Pethick}
\begin{gather}
	D_{\pm k} = u_k\beta_{\pm k} - v_k\beta_{\mp k}^\dagger, \label{eq:Bogoliubov}\\
	u_k,v_k\in\mathbb{R},\quad u_k^2-v_k^2=1,\quad [\beta_{\pm k},\beta_{\pm k'}^\dagger]=\delta_{kk'}, \nonumber
\end{gather}
which again preserve commutation relations for the new Bogoliubov bosons.  After the transformation, $\mathcal{H}_Q$ takes the form
\begin{gather}
  \mathcal{H}_Q=L_0+\sum_k{\left[\varepsilon(k)\beta_k^\dagger\beta_k+\frac12\varepsilon(k)-\frac12\varepsilon_0(k)\right]},\label{eq:NN-HQB}\\
	\varepsilon(k)=\sqrt{\varepsilon_0^2(k)-\varepsilon_1^2(k)},\nonumber\\
	\varepsilon_0(k)=L_1-2L_3\cos k,\quad \varepsilon_1(k)=2L_2+2L_3\cos k,\nonumber
\end{gather}
where we returned to the summation over all the wavenumbers. As the factors $L_1$ to $L_3$ do not depend on $N$, excitation energies $\varepsilon(k)$ are independent of it, too.

We note here for later reference that the products of Bogoliubov coefficients $v_k^2$ and $u_k v_k$ can be expressed in terms of $\varepsilon(k)$, $\varepsilon_0(k)$, and $\varepsilon_1(k)$ as
\begin{equation}
      v_k^2 = \frac12\left[\frac{\varepsilon_0(k)}{\varepsilon(k)}-1\right]\quad\text{and}\quad
      u_k v_k = \frac{\varepsilon_1(k)}{2\varepsilon(k)}.
      \label{eq:ukvk}
\end{equation}

\begin{figure}
	\includegraphics[width=0.9\columnwidth]{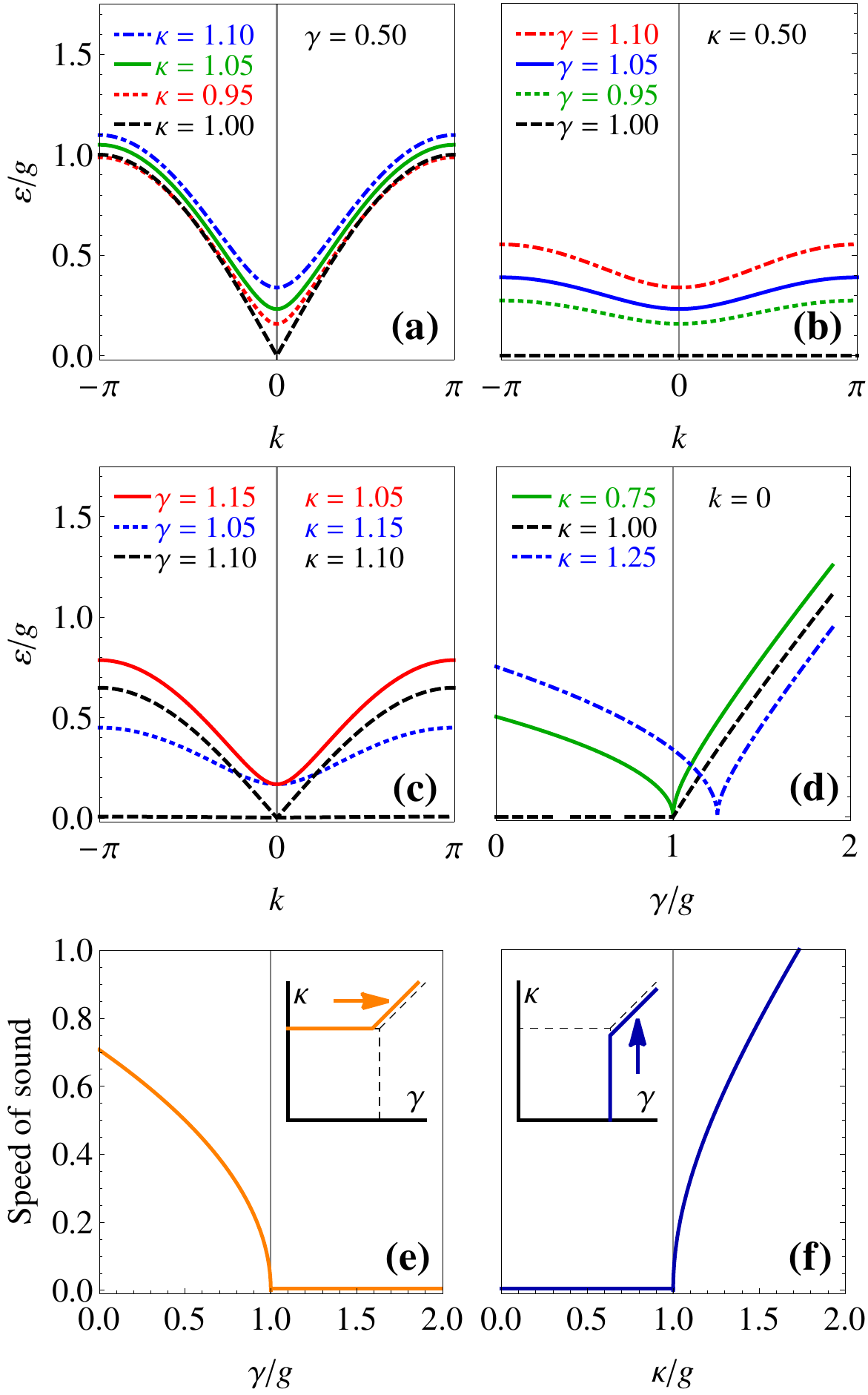}
	\caption{\label{fig:NN-dEG}(Color online) The excitation energy dispersion $\varepsilon(k)$ across the (I--II) boundary~(a), across the (I--III) boundary~(b), and across the (II--III) boundary~(c) as well as the energy gap $\varepsilon(k=0)$ as a function of $\gamma$~(d).  Panels (e) and (f) show the speed of sound $(\dd\varepsilon/\dd k)_{k=0}$ along the phase boundaries.}
\end{figure}

Analysis of \eqref{eq:NN-HQB} shows that the ground-state excitation energy is minimal at $k=0$ and varies with $\kappa$ and $\gamma$ parameters as shown in Fig.~\ref{fig:NN-dEG}d.  The energy dispersion is quadratic and gapped in the vicinity of $k=0$ for parameters away from critical lines.  At the phase boundary (I--II) the gap closes with a linear dispersion (see Fig.~\ref{fig:NN-dEG}a), denoting, much like in the Ising model, the transition from an unordered paramagnetic phase to the ferromagnetic one.  In this case, the softening of the collective excitation leads to long-range correlations resembling the Ising critical point in quantum magnetism \cite{2014-Acevedo-PRL,2012-Schiro-PRL}.

At the boundary (I--III), on the other hand, energy gap becomes zero at all the wavenumbers (see Fig.~\ref{fig:NN-dEG}b) thus allowing for the collective excitations of any wavelength and marking the two phases between which the transition occurs as lacking long-range ordering.

At the boundary (II--III) the form of dispersion relation changes drastically: Approaching the boundary from within the phase (II), the gap closes with the linear dispersion in the limit $\gamma\rightarrow\kappa$,  while when approaching the boundary from within the phase (III), all the modes are gapless in the limit $\kappa\rightarrow\gamma$ (Fig.~\ref{fig:NN-dEG}c).  This jump in the form of the gap closing indicates the first-order QPT.

The linear dispersion around $k=0$ at the phase boundaries is phonon-like and lets us define the group velocity $c=(\dd\varepsilon/\dd k)_{k=0}$.  The way it changes with parameters is shown in Figs.~\ref{fig:NN-dEG}e and \ref{fig:NN-dEG}f. At the phase boundary (II--III) the velocity of propagation exhibits a discontinuous behavior, which is another signature of the first-order QPT.

Now, to obtain the ground state of the Hamiltonian \eqref{eq:NN-LMG} we rely on the diagonalized Hamiltonian \eqref{eq:NN-HQB}, from which one can see that the ground state $|G\rangle$ is defined by the condition $\beta_k^\dagger\beta_k|G\rangle=0$, which leads to the expression
\begin{equation}
  |G\rangle=\bigotimes_{k>0}\mathcal{S}\left(\chi_k\right)\,\tilde{\mathcal{D}}\left(\alpha_k\sqrt{j}\right)\,\tilde{\mathcal{D}}\left(\alpha_{-k}\sqrt{j}\right)|0_{k},0_{-k}\rangle,
  \label{eq:BogoliubovVacuum}
\end{equation}
where $\alpha_{\pm k}=\frac{1}{\sqrt{N}}\sum_{l=1}^N{\alpha_l \ee^{\mp\ii k l}}$ are the Fourier images of the mean fields and $D_{\pm k}|0_{k},0_{-k}\rangle=0$.  We also have used the displacement operators in the Fourier space \cite{1996-Hu-PRL}  
\begin{equation}
  \tilde{\mathcal{D}}\left(\alpha_{\pm k}\sqrt{j}\right)=\exp\left[\left(\alpha_{\pm k} D_{\pm k}^{\dagger}-\alpha_{\pm k}^* D_{\pm k}\right)\sqrt{j}\right].
  \label{eq:DispOpQuansimomentum}
\end{equation}
Similarly to Ref.~\onlinecite{2013-Zhao-PRA}, the ground state is a product of two-mode squeezed states with squeezing parameters $\chi_k=\artanh[\varepsilon_1(k)/\varepsilon_0(k)]$, where $\varepsilon_0(k)$ and $\varepsilon_1(k)$ are defined in \eqref{eq:NN-HQB}.
The two-mode squeezing operator \cite{1996-Hu-PRL,2013-Zhao-PRA} is
\begin{equation}
  \mathcal{S}(\chi_k)=\exp\left[\chi_k\left(D_{-k}D_k- D^{\dagger}_{-k} D^{\dagger}_k\right)\right].
  \label{eq:TwoModeSqueezing}
\end{equation}

\section{Correlation functions}\label{sec:correlation}
In order to further characterize the phases of the system, it is useful to calculate correlations of some observables between different sites in the ground state for each of the phases.  One of the obvious choices is to consider correlations between components of total angular momenta of some site $l$ and a site $l+r$, which is $r$ bonds away from the former.  Thus we are interested in functions
\begin{equation}
	C_{\xi\xi'}(r)=\frac{1}{2jN}\sum^{N}_{l=1}\langle J_l^\xi J_{l+r}^{\xi'}\rangle_{G},
      \label{eq:DefCorrelation}
\end{equation}
where $\xi,\xi'\in\{x,y,z\}$.  To simplify the notation we have defined the expectation value of an operator $\mathcal{O}$ in the ground state as $\langle \mathcal{O}\rangle_{G}=\langle G|\mathcal{O}|G\rangle$.  The scaling factor $\frac1{2j}$ is introduced to maintain consistency with the original Hamiltonian \eqref{eq:NN-LMG}.  We shall now approach the calculation of correlation functions first in limit cases (see Sec.~\ref{sec:symmetries}) and then taking into account the equal-mean-field ansatz.

\subsection{Limit cases}\label{sec:CorLimit}
As in Sec.~\ref{sec:symmetries}, we can look at three limit cases for an arbitrary value of the total angular momentum $j$.  If $g\gg \gamma,\kappa$, the ground state of the system is $|G^z_{1,1,\ldots,1}\rangle$ [see \eqref{eq:GStates} for notation].  The correlation functions in this state are
\begin{gather}
	\frac{1}{2j}\langle G^z_{1,1,\ldots,1}|J_l^z J_{l+r}^z|G^z_{1,1,\ldots,1}\rangle = \frac{j}2, \\
	\langle G^z|J_l^x J_{l+r}^x|G^z\rangle = \langle G^z|J_l^y J_{l+r}^y|G^z\rangle = 0. \nonumber
\end{gather}

In the limit $\kappa\gg g,\gamma$, the ground state is twofold degenerate, so in order to take both the states into account, we consider a symmetric combination $|G\rangle=\frac{1}{\sqrt 2}\left(|G^y_{1,1,\ldots,1}\rangle+G^y_{0,0,\ldots,0}\rangle\right)$ leading to correlation functions
\begin{gather}
	\frac{1}{2j}\langle G|J_l^y J_{l+r}^y|G\rangle = \frac{1}{4j}\left(j^2+j^2\right) = \frac{j}{2},\\
	\langle G|J_l^x J_{l+r}^x|G\rangle = \langle G|J_l^z J_{l+r}^z|G\rangle = 0. \nonumber
\end{gather}

\begin{table*}
	\caption{\label{tab:M}Classical correlations $jM_g$ and parameters $M_0$ through $M_3$ used in microscopic correlations \eqref{eq:NN-Cor} in regions with ground-state mean fields $\alpha_l=\alpha_\mathrm{cr}$ between angular momenta components $J_i^\xi$ and $J_{i+r}^\xi$.  The ``$\pm$'' column shows whether in \eqref{eq:NN-Cor} the upper ($+$) or the lower ($-$) sign should be chosen.}
	\begingroup
	\renewcommand{\arraystretch}{1.2}
	\setlength\arraycolsep{8pt}
	\[\begin{array}{ccccccccc}
		\hline\hline
		\text{Region} & \alpha_\mathrm{cr}& \xi & jM_g & M_0 & M_1 & M_2 & M_3 & \pm\\
		\hline
		&                                 & x & 0 & 0 & 0 & 0 & \frac14 & + \\
		\text{I} & 0                      & y & 0 & 0 & 0 & 0 & \frac14 & - \\
		&                                 & z & \frac{j}{2} & 0 & -\frac12 & 0 & 0 & +\\
		\hline
		&                                 & x & 0 & 0 & 0 & 0 & -\frac{g-\kappa}{8\kappa} & + \\
		\text{II} & \ii\sqrt{1-g/\kappa}  & y & \frac{j}{2}\left(1-\frac{g^2}{\kappa^2}\right) & \frac{(g+\kappa)^2}{8\kappa(g-\kappa)} & \frac{5\kappa^2+2g\kappa-3g^2}{8\kappa(g-\kappa)} & -\frac{3\kappa^2+2g\kappa-g^2}{16\kappa(g-\kappa)} & \frac{g^2}{2\kappa(g-\kappa)} & - \\
		&                                 & z & \frac{j}{2}\frac{g^2}{\kappa^2} & 0 & \frac{g}{2\kappa} & 0 & -\frac{g+\kappa}{2\kappa} & - \\
		\hline
            &                                 & x & \frac{j}{2}\left(1-\frac{g^2}{\gamma^2}\right) & \frac{(g+\gamma)^2}{8\gamma(g-\gamma)} & \frac{5\gamma^2+2g\gamma-3g^2}{8\gamma(g-\gamma)} & \frac{3\gamma^2+2g\gamma-g^2}{16\gamma(g-\gamma)} & -\frac{g^2}{2\gamma(g-\gamma)} & + \\
            \text{III} & \sqrt{1-g/\gamma}    & y & 0 & 0 & 0 & 0 & \frac{g-\gamma}{8\gamma} & - \\
            &                                 & z & \frac{j}{2}\frac{g^2}{\gamma^2} & 0 & \frac{g}{2\gamma} & 0 & \frac{g+\gamma}{2\gamma} & + \\
            \hline\hline
	\end{array}\]
	\endgroup
\end{table*}

In the third limit, $\gamma\gg g,\kappa$, the ground state is $2^N$-fold degenerate with different sites having spin projections either $j$ or $-j$ independently of one another.  This limit implies that for a small angular momentum $j$ there is tunneling between different ground states, making all the states to have equal probabilities.  So, as in the large-$\kappa$ limit, we consider the ground state to be an equally weighted combination of  ground states
\begin{equation}
	|G\rangle = \frac{1}{\sqrt{2^N}}\sum_{p_1,p_2,\ldots,p_N}{|G^x_{p_1,p_2,\ldots,p_N}\rangle}
  \label{eq:gammaGroundState}
\end{equation}
with $p_i$ taking values 0 and 1.  The $z$--$z$ and $y$--$y$ correlations are zero (cf. previous limits), as well as the $x$--$x$ correlations
\begin{equation}
	\langle J_l^x J_{l+r}^x\rangle = \frac{1}{2^{N+1}jN}\sum_l\sum_{p_1,p_2,\ldots,p_N}(-1)^{p_l}(-1)^{p_{l+r}}j^2=0
	\ . 
\end{equation}
This occurs independent of $r$, as the pure states $|G^x_{p_1,p_2,\ldots,p_N}\rangle$ are orthogonal and $p_l$ (as well as $p_{l+r}$) is zero $2^{N-1}$ times and one another $2^{N-1}$ times, making the positive terms appear in the sum precisely the same number of times as the negative ones.

It thus may be concluded that in the large-$\gamma$ limit the system exhibits no correlations whatsoever in any of the components [at least when it is in the state \eqref{eq:gammaGroundState}], while large-$\kappa$ and large-$g$ limits show correlations in $y$- and $z$-\nobreakdash directions, respectively.

\subsection{Equal-mean-field case}\label{sec:CorEqual}
In this section we consider correlation functions in a ground state that has full translational invariance.  It is important to note that both the paramagnetic and the ferromagnetic ground states inherit this property.  In the region (III), however, due to the exponential degeneracy, there are only two ground states that possess full translational invariance.  Other ground states have lower translational symmetry.

Working in the equal-mean-field ansatz we can use the Bogoliubov vacuum \eqref{eq:BogoliubovVacuum} as the ground state $|G\rangle$.  $C_{\xi \xi'}$ is a sum of correlations of classical background, which are of order $j$, and correlations of microscopic fluctuations of angular momenta, which are of order 1, much like the Hamiltonian \eqref{eq:NN-HQ} consists of a classical ground-state energy $jE_g\sim O(j)$ and microscopic fluctuations $\mathcal{H}_Q\sim O(1)$ thereupon. (As we consider only minima of $E_g$, linear terms vanish.)

By calculating both classical and microscopic parts of correlation functions in the limit $r\to\infty$ (implying $N\to\infty$), we can classify the phases according to the long-range ordering.  We do it by applying Holstein--Primakoff transformations to the component products followed by Fourier transformations thereof.  After these operations we get the following correlations between similar components [restricting wavenumbers to positive values of $k$ again, cf. \eqref{eq:NN-HQFP}]:
\begin{align}
	C_{\xi\xi}(r) &= jM_g + M_0 \nonumber\\
	&\quad + \sum_{k>0}{\left(M_1\pm 2M_3\cos kr\right)\langle D_k^\dagger D_k + D_{-k}^\dagger D_{-k}\rangle_G} \nonumber\\
	&\quad + \sum_{k>0}{\left(2M_2+2M_3\cos kr\right)\langle D_k D_{-k}+D_k^\dagger D_{-k}^\dagger \rangle_G}
  \label{eq:NN-Cor-beta}
\end{align}
with parameters $M_g$ and $M_0$ through $M_3$ as well as the sign in the first sum defined in Tab.~\ref{tab:M}.  Note that approaching the limit cases where either of the parameters $g$ or $\kappa$ is much larger than the others, the macroscopic part of correlation functions is in accordance with the functions from the previous section, as in the respective regions the ground state consists of parallel spins.  The value of the $C_{xx}$ in the region (III), on the other hand, is different, which is due to the exponential degeneracy of the ground state.  If in Sec.~\ref{sec:CorLimit}, a small total angular momentum $j$ is considered, the tunneling between different states is highly favorable, making the resulting ground state uncorrelated.  The equal-mean-field ansatz for large $j$, though, implies that once initialized in a state with all the spins parallel, the system stays in that state and no tunneling occurs resulting in long-range correlations.

Fourier-transformed correlation functions \eqref{eq:NN-Cor-beta} can then be mapped onto Bogoliubov bosons $\beta_k,\beta^{\dagger}_k$ obtained from diagonalization of the original Hamiltonian taking into account expressions \eqref{eq:ukvk} as well as that $\langle\beta_k^\dagger\beta_{k}\rangle_{G}=\langle\beta_{-k}^\dagger\beta_{-k}\rangle_{G}=\langle\beta_k\beta_{-k}\rangle_{G}=\langle\beta_k^\dagger\beta_{-k}^\dagger\rangle_{G}=0$ in the ground state \eqref{eq:BogoliubovVacuum}.

For a finite size $N$ of the ring, the final expression for the correlation function \eqref{eq:NN-Cor-beta} reads
\begin{align}
	C_{\xi\xi}(r) &= jM_g+ M_0+\frac{2}{N}M_1\sum_{k>0}\left[\frac{\varepsilon_0(k)}{\varepsilon(k)}-1\right]\nonumber\\
	&\qquad\quad-\frac{4}{N}M_2\sum_{k>0}\frac{\varepsilon_1(k)}{\varepsilon(k)} \nonumber \\
    &\qquad\quad\pm\frac{2}{N}M_3\sum_{k>0}{\cos kr\left[\frac{\varepsilon_0(k)\mp\varepsilon_1(k)}{\varepsilon(k)}-1\right]},
	\label{eq:NN-Cor}
\end{align}
where we used \eqref{eq:ukvk} to express Bogoliubov coefficients in terms of $\varepsilon(k)$, $\varepsilon_0(k)$, and $\varepsilon_1(k)$.

In the limit $N\to\infty$, we can rewrite the sums as integrals, assuming $\dd k\equiv\frac{2\pi}{N}$ to get
\begin{align}
	C_{\xi\xi}(r) &= jM_g+ M_0+\frac{M_1}{\pi}\int_0^\pi{\left[\frac{\varepsilon_0(k)}{\varepsilon(k)}-1\right]\,\dd k} \nonumber\\
	&\qquad\quad-\frac{2M_2}{\pi}\int_0^\pi{\frac{\varepsilon_1(k)}{\varepsilon(k)}\,\dd k} \nonumber \\
    &\qquad\quad\pm\frac{M_3}{\pi}\int_0^\pi{\left[\frac{\varepsilon_0(k)\mp\varepsilon_1(k)}{\varepsilon(k)}-1\right]\cos kr\,\dd k}.
	\label{eq:NN-Cor-Int}
\end{align}

The macroscopic part of \eqref{eq:NN-Cor-Int}, namely $jM_g$, is proportional to the respective projection of any of the angular momenta [$C^\mathrm{cl}_{\xi\xi}=\frac{1}{2j}(J_l^\xi)^2$], as we initialize the ring in the ground state where all the angular momenta of different sites are parallel and macroscopically static.  Thus $C_{\xi\xi}^\mathrm{cl}$ is maximized. 

The microscopic correlations, too, may be separated into two parts: the background, which is independent of the distance $r$
\begin{equation}
	C_{\xi\xi}^\infty=\frac1\pi\int_0^\pi{\left\{M_0+M_1\left[\frac{\varepsilon_0(k)}{\varepsilon(k)}-1\right]-2M_2\frac{\varepsilon_1(k)}{\varepsilon(k)}\right\}\,\dd k},
	\label{eq:NN-Cor-Inf}
\end{equation}
and oscillations around this background, decaying with $r$ 
\begin{equation}
	C_{\xi\xi}^\mathrm{osc}(r)=\pm\frac{M_3}\pi\int_0^\pi{\left[\frac{\varepsilon_0(k)\mp\varepsilon_1(k)}{\varepsilon(k)}-1\right]\cos kr\,\dd k.}
	\label{eq:NN-Cor-Osc}
\end{equation}

\begin{figure}
	\includegraphics[width=0.8\columnwidth]{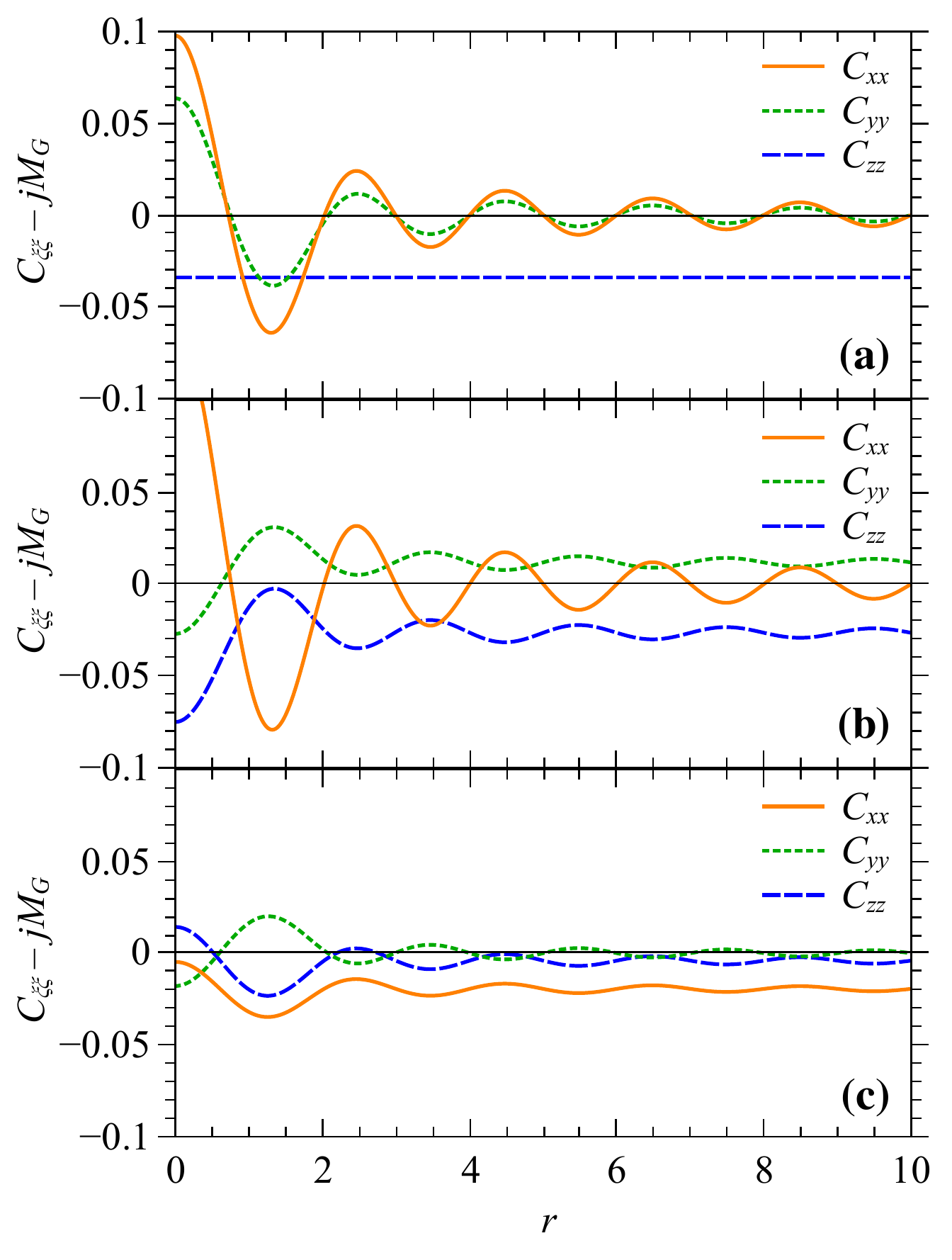}
	\caption{\label{fig:NN-CorOsc}(Color online) Microscopic parts of correlation functions $C_{\xi \xi}(r)-j M_g$ in the regions (I) with $\gamma=0.5$ and $\kappa=0.5$~(a), (II) with $\gamma=0.5$ and $\kappa=1.5$~(b), and (III) with $\gamma=1.5$ and $\kappa=0.5$~(c).}
\end{figure}
In the limit $r\to\infty$ the oscillating part vanishes leaving us with $C_{\xi\xi}(\infty)-jM_g=C_{\xi\xi}^\infty$.  Microscopic parts of the correlation functions are plotted in Fig.~\ref{fig:NN-CorOsc}.  The plots show that they differ distinctly in different regions only in the background terms $C_{\xi\xi}^\infty$ and the amplitude of the oscillations.  Both the quasiperiod and the damping remain the same for all $\gamma$ and $\kappa$.

In the region (I), only $z$--$z$ correlations persist in the long-range limit (Fig.~\ref{fig:NN-CorOsc}a), repeating the behavior of the macroscopic correlations.  The same correspondence between the macroscopic and microscopic correlations is also found in the regions (II) and (III).  That said, we can classify the phases limiting ourselves to the macroscopic functions without loss of generality to get that the long-range ordering exists in all the phases in corresponding spin components, i.e. in $J_z$'s in region (I), $J_y$'s in region (II), and $J_x$'s in region (III).  Note again that the presence of the long-range ordering in the region (III) is subject to tunneling possibility between the ground states and initial conditions.

\section{Conclusions}
We have discussed the quantum phase transitions in a network of LMG systems coupled via ferromagnetic interactions.  In the particular case of a ring topology of the network, we have shown that the ground state can be obtained by calculating the quantum corrections about the mean-field solution.  Within the mean-field approach, the ground state can be interpreted as an alignment of angular momenta of the individual sites along the directions that minimize the classical energy of the network.  The phase diagram determined from this assumption shows three distinct phases in $\gamma\kappa$\nobreakdash-space in the thermodynamic limit. Such a phase diagram can also be obtained from observing the quantum corrections, i.e., the energy of the collective excitations above the ground state, because the system is gapless along the critical lines \cite{2011-Sachdev}. 

In the particular case of a ground state of the network with full translational invariance, the correlation functions between angular momentum components are clearly distinct for different phases, showing $\kappa$- and $\gamma$-dominated phases being ferromagnetic in $y$- and $x$\nobreakdash-directions, respectively, and $g$-dominated phase showing no long-range ordering.  The order parameter increases with the strength of exchange interaction $\kappa$ and decreases with the strength of local interaction $\gamma$.

Apart from the  possibility of experimental realization of coupled LMG models by using BEC in optical lattices \cite{2010-Gross-Nature,2010-Zibold-PRL,2006-Morsch-RevModPhys,2010-Theocharis-PRA}, our model could be realized in single-molecule and single-chain magnets, as well as in nanomagnets \cite{2011-Chotorlishvili-PRB,2011-Konstantinidis-JPhys,2004-Mikeska-Springer,2006-Coulon,2008-Wernsdorfer-PRL}. The method we have developed in this work can be extended to study other kinds of networks consisting of coupled mean-field-type critical systems, e.g., the Dicke models \cite{2003-Emary-PRL} and spinor Bose gases within the single-mode approximation \cite{2013-Stamper-RevModPhys}.

Further studies may include the detailed description of the first-order phase transition at the boundary between regions (II) and (III) and what happens with correlation functions in its vicinity.  In this paper we have not considered the issue of antiferromagnetic coupling, i.e., the case when $\kappa_{ll'}\leqslant 0$.  In this context, it would be interesting to explore other topologies of the network, to study the emergence of frustration and exotic states such as spin ice \cite{2010-Balents-Nature}.

\acknowledgments{%
  The authors are thankful to Gernot Schaller for fruitful discussions and gratefully acknowledge financial support through grants BRA~1528/7, BRA~1528/8, BRA~1528/9, SFB~910 (VMB and TB), and through DAAD scholarship A/12/84446 (AVS).%
}

\bibliographystyle{apsrev4-1}
\bibliography{Xbib}

\end{document}